\documentclass[twocolumn,showpacs,epsfig,fleqn,nobibnotes]{revtex4}

\usepackage{epsfig}
\begin{document}

\newcommand{\qs}{Q_{\rm sat}}
\newcommand{\qsa}{Q_{\rm sat, A}}
\newcommand{\rr}{\mbox{\boldmath $r$}}
\newcommand{\rrn}{\mbox{$r$}} 
\newcommand{\rp}{\mbox{\boldmath $p$}} 
\newcommand{\rqq}{\mbox{\boldmath $q$}} 
\newcommand{\lsim}{\raisebox{-0.5mm}{$\stackrel{<}{\scriptstyle{\sim}}$}}
\newcommand{\gsim}{\raisebox{-0.5mm}{$\stackrel{>}{\scriptstyle{\sim}}$}}
\def\simge{\mathrel{%
   \rlap{\raise 0.511ex \hbox{$>$}}{\lower 0.511ex \hbox{$\sim$}}}}
\def\simle{\mathrel{
   \rlap{\raise 0.511ex \hbox{$<$}}{\lower 0.511ex \hbox{$\sim$}}}}

\title{Geometric scaling in ultrahigh energy neutrinos and nonlinear perturbative QCD}
\author{Magno V.T. Machado$^{\,\star,\,\ddag}$}

\affiliation{${}^{\star}$Departamento de F\'{\i}sica de Part\'{\i}culas. 15706 Universidade de Santiago de Compostela, Spain\\
${}^{\ddag}$Universidade Estadual do Rio Grande do Sul, Unidade de Bento Gon\c{c}alves. CEP 95700-000, Brazil}


\begin{abstract}
It is shown that in ultrahigh energy inelastic neutrino-nucleon(nucleus) scattering  the cross sections for the boson-hadron(nucleus) reactions should exhibit  geometric scaling on the single variable $\tau_A =
Q^2/Q_{\mathrm{sat,A}}^2$. The dependence on energy and atomic number of the charged/neutral current cross sections are encoded in the saturation momentum $Q_{\mathrm{sat,A}}$. This fact  allows an analytical computation of the neutrino scattering on nucleon/nucleus at high energies, providing a theoretical parameterization based on the scaling property. 
\end{abstract}
\maketitle

\section{Introduction}

One important property of the  nonlinear perturbative QCD  approaches for high energy deep inelastic $ep(A)$ scattering   is
the prediction of the  geometric scaling. Namely,  the total
$\gamma^* p(A)$ cross section at large energies is not a function of
the two independent variables $x$ and $Q^2$, but is rather a
function of the single variable $\tau_A = Q^2/Q_{\mathrm{sat,A}}^2$. As usual, $Q^2$ is the photon virtuality and $x$ the Bjorken variable. It was demonstrated \cite{travwaves} that geometric scaling is the exact asymptotic solution of a general class of nonlinear evolution equations \cite{KPP,BK}  and it appears as a universal property of these kind of equations. The specific scaling solutions correspond to traveling wave solutions of those equations. The saturation momentum $Q_{\mathrm{sat,A}}^2(x;\,A)\propto\frac{xG_A(x,\, Q_{\mathrm{sat}}^2)}{\pi R_A^2} \simeq A^{\,\alpha}\,x^{-\lambda}$ ($\alpha \simeq 1/3$, $\lambda \simeq 0.3$)  is connected with the phenomenon of gluon saturation. It is expected that the rise of the gluon distribution function at small values of Bjorken $x$ be naturally tamed by saturation and circumvent the subsequent violation of unitarity. In principle, geometric scaling is predicted to be present only on process dominated by low momenta. However, it was theoretically demonstrated \cite{Iancu:2002tr} that the geometric scaling is preserved by the QCD evolution  up
to relatively large virtualities, within the kinematical window $Q_{\mathrm{sat}}^2 (x) \simle Q^2 \simle Q_{\mathrm{sat}}^4 (x)/\Lambda^2_{\mathrm{QCD}}$. That is, the scaling property extends towards very large virtualities provided one stays in low-$x$. This kinematical window is further enlarged due to the nuclear enhancement of the saturation scale. In  Ref. \cite{Stasto:2000er}, the observation that the DESY-HERA $ep$ collider data
on the proton structure function $F_2$ are consistent with scaling
at $x \leq 0.01$ and $Q^2 \leq 400$ GeV$^2$ was for the first time presented. Similar behavior was further observed on  electron-nuclei processes \cite{Freund:2002ux}. Geometric scaling has been also observed in inclusive charm production \cite{magvicprl} and predicted to be present also in heavy quark production in lepton-nuclei scattering \cite{gomampla}.

Recently, the high energy lepton-hadron, proton-nucleus and nucleus-nucleus collisions have been related through geometric scaling \cite{Armesto_scal}. Within the color dipole picture and making use of a rescaling of the impact parameter of the $\gamma^*h$ cross section in terms of hadronic target radius $R_h$, the nuclear dependence of the $\gamma^*A$ cross section is absorbed in the $A$-dependence of the saturation scale via geometric scaling. The relation reads as
\begin{eqnarray}
& &  \sigma^{\gamma^*A}_{tot}\,(\tau_A)  =  \frac{\pi R_A^2\,\sigma^{\gamma^*p}_{tot}\,(\tau_p=\tau_A)}{\pi R_p^2},\label{sig_nuc_scal}\\
 & &  \tau_A  = \tau_p\left(\frac{ \pi R_A^2}{A 
                       \pi R_p^2}\right)^{\Delta}, \hspace{0.5cm} \tau_p = \frac{Q^ 2}{Q_{\mathrm{sat}}^2(x)},
\label{nuclear_scaling}
\end{eqnarray}
where the nuclear saturation scale was assumed to rise with the quotient of the transverse parton densities to the power $\Delta=1/\delta$. The nucleon saturation momentum is set to be $Q^2_{\mathrm{sat}}(x)=(x_0/\bar{x})^{\lambda}$  GeV$^2$, where $x_0= 3.04\cdot 10^{-4}$, $\lambda=0.288$ and $\bar{x}=x\,[1+ (4m_f^2/Q^2)]$, with $m_f=0.14$ GeV, as taken from \cite{GBW}. The nuclear radius is given by $R_A=(1.12 A^{1/3}-0.86 A^{-1/3})$ fm. The following functional shape (scaling curve) for the photoabsortion cross section has been considered based on theoretical motivations \cite{Armesto_scal}:
\begin{eqnarray}
  \sigma^{\gamma^* p}_{tot}\,(\tau_p) = \bar{\sigma}_0\,
  \left[ \gamma_E + \Gamma\left(0,\beta \right) +
         \ln \left(\beta\right) \right],\hspace{0.5cm} \beta = a/\tau_p^{b}\,,
       \label{sigtot_param_tau}
\end{eqnarray}
where $\gamma_E$ is the Euler constant and $\Gamma\left(0,\beta\right)$
the incomplete Gamma function. The parameters for the proton (nucleon) case were obtained from a fit to the small-$x$ $ep$ DESY-HERA data, producing $a=1.868$, $b=0.746$ and the overall  normalization was fixed by $\bar\sigma_0=40.56$ $\mu$b. The parameters for the nuclear saturation scale were determined by fitting the available lepton-hadron data using the relation in Eq. (\ref{sig_nuc_scal}) and the same scaling function, Eq. (\ref{sigtot_param_tau}). One obtains $\delta=0.79\pm0.02$ and $\pi R_p^2=1.55 \pm 0.02$ fm$^2$. The value of $\delta$ favors a nuclear dependence $Q_{\mathrm{sat,A}}^2\propto A^{4/9}\, Q_{\mathrm{sat}}^2$, which it implies a faster increasing in comparison with the usual ansatz $Q_{\mathrm{sat,A}}^2 \propto A^{1/3}$.

The theoretical and phenomenological achievements summarized above have direct consequences on the computation of the cross sections of interaction of neutrinos with nucleons and atomic nuclei. Ultrahigh energy (UHE) neutrino reactions are sensitive upon the behavior of the nucleon/nuclear structure functions at extremely low-$x$ values and relatively large scales of electroweak boson virtualities. Namely, they can probe the domain $x\simeq m_{W,Z}^2/E_{\nu} \sim 10^{-8}$ at $E_{\nu}\sim 10^{12}$ GeV and virtualities $Q^2\sim m_{W,Z}^2\approx 10^{4}$ GeV$^2$, where $m_{W,Z}$ are the boson masses. Therefore, accurate predictions for UHE neutrinos require  precise extrapolation of the structure functions to the small-$x$.  Recently, the approaches containing saturation effects and/or scaling property have been compared and their main features were identified and investigated \cite{nu_prd,KUTAK}.  A very important feature is that, within the color dipole framework \cite{DIPOLEPIC}, the charged (CC) and neutral (NC) current structure functions are described by the same mathematical expressions as the proton structure function up to a different coupling of the electroweak bosons. Therefore, the geometric scaling property should be present also in neutrino scattering on hadron targets and allows to obtain the dependences on energy and atomic number of CC/NC cross sections, which are encoded in the nuclear saturation momentum. In fact, geometric scaling holds for a large region of phase space and thus  it contributes for an important part of the integrated cross section. Furthermore, the upper limit for scaling window should be enhanced by a factor $\sim 10-100$ in the nuclear case. In what follows, the weak boson-hadron/nucleus cross section is computed and it is shown to exhibit geometric scaling on the scaling variable $\tau_A$. The CC and NC structure function are further studied and compared with the available high energy approaches. The scaling property will allow to obtain simple analytical expressions for the CC/NC neutrino cross sections at high energies. They  will be used to provide theoretical parameterizations for the UHE neutrino scattering on nucleons and nuclei.

\section{UHE neutrino cross section}

Deep inelastic neutrino scattering can proceed via  $W^{\pm}$ (charged current interactions - CC) or $Z^0$ (neutral current interactions - NC) exchanges.  The standard kinematical variables describing them are given by $s=2\,m_N E_{\nu}$ (center-of-mass energy squared), $Q^2$ (boson virtuality), Bjorken $x$ and $y=Q^2/xs$ (inelasticity variable). Here, $m_N$ is the nucleon mass and $E_{\nu}$ labels the neutrino energy. At small-$x$, a successful framework describing QCD interactions is provided by the color dipole formalism \cite{DIPOLEPIC}, which allows an all-twist computation of the structure functions. The physical picture of the interaction is the deep inelastic scattering (DIS) at low $x$  viewed as the result of the interaction of a color $q \bar{q}$ dipoles, which are fluctuations of the electroweak gauge bosons, with the hadron target. The interaction is modeled via the dipole-target cross section, whereas the boson fluctuation in a color dipole is given by the corresponding wave function. The DIS structure functions for neutrino scattering read as \cite{nu_prd},
\begin{eqnarray}
& & F_{T,L}^{\mathrm{CC,NC}}(x,Q^2)  =  \frac{Q^2}{4\,\pi^2}\, \sigma_{tot}\,(W^{\pm}(Z^0)\, N \rightarrow X)\,,\label{weak_sfs}\\
& & \sigma_{tot}\,(W^{\pm}(Z^0) \,N)  = \int d^2 \rr \,\int_0^1 dz \,
| \psi^{W^{\pm},Z^0}\,|^2\,\sigma_{dip}\,, 
\label{sig_boson}
\end{eqnarray}
where  $\rr$ denotes the transverse size of the color  dipole, $z$ the 
longitudinal momentum fraction carried by a quark and  $\psi^{W,Z}_{T,L}$ are the  sum over dipoles of the wave functions of the charged or neutral gauge bosons, respectively. Their explicit expressions can be found for instance in Refs. \cite{nu_prd,KUTAK,JAMAL}. Here one considers  only four flavors ($u,d,s,c$) assumed to be massless. Heavy quarks $(b,t)$ give relatively small contribution and will be disregarded. Color dipoles  contributing  to Cabibbo favored transitions are  $ u \bar d \, (d \bar u)$,  
$ c \bar s \,(s \bar c)$ for CC interactions  and $ u \bar u, d \bar d , s \bar s, c \bar c $ for NC interactions. The dipole cross section $\sigma_{dip}\,(x,\rr;A)$, describing the dipole-target interaction, is substantially affected by saturation effects at dipole sizes $\rr \simge 1/Q_{\mathrm{sat}}$.

Based on the fact that the expressions for the photon wavefunction and the electroweak gauge bosons, appearing on Eq. (\ref{sig_boson}),  are exactly the same up to the different coupling of the bosons to the quark color dipole, we can simply write,
\begin{eqnarray}
\sigma_{tot}^{(W^{\pm}N)}\,(x,\,Q^2;\,A)  = \frac{4}{\alpha_{\mathrm{em}}\sum_{f} e_f^2}\,\sigma_{tot}^{(\gamma^*N)}\,(x,\,Q^2;\,A)\,,\label{sigboscc}\\ 
\sigma_{tot}^{(Z^0N)}\,(x,\,Q^2;\,A)  = \frac{K_{\mathrm{chiral}}}{\alpha_{\mathrm{em}}\sum_{f} e_f^2}\,\sigma_{tot}^{(\gamma^*N)}\,(x,\,Q^2;\,A)\label{sigbosnc} \,,
\end{eqnarray}
where $\alpha_{\mathrm{em}}$ is the QED constant coupling, $e_f$ is the electric charge of the quark of flavor $f$. The constant $K_{\mathrm{quiral}}=(L_u^2+L_d^2+R_u^2+R_d^2)$ is the sum of the chiral couplings expressed as functions of the Weinberg angle $\theta_W$.


\begin{figure}[t]
\includegraphics[scale=0.42]{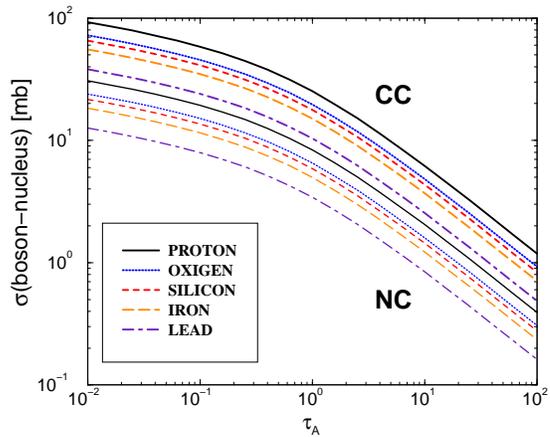}
\caption{Geometric scaling for the electroweak boson-nucleus cross section as a function of the scaling variable $\tau_A$ for different atomic numbers (normalized to nucleon).}
\label{fig:1}
\end{figure}

The scaling present in the lepton-hadron cross section at high energies, as quantified by Eq. (\ref{sig_nuc_scal}) and further experimentally demonstrated,  is automatically translated to the UHE neutrino scattering on nucleon or nucleus.  In Fig. \ref{fig:1} one presents the boson-hadron cross sections, Eqs. (\ref{sigboscc}-\ref{sigbosnc}), as a function of the scaling variable $\tau_A$ for distinct nucleus as well as for the nucleon. They are normalized to the nucleon. The cross sections  exhibit geometric scaling, verifying a transition in the
behavior on $\tau_A$  of the cross section from a smooth dependence
at small $\tau_A$ and an approximated $1/\tau$ behavior at large
$\tau_A$. Similarly to the lepton-hadron case, the transition point is placed at $\tau_A =1$. The asymptotic $1/\tau^b$
dependence reflects the fact that the cross section
scales as $Q_{\mathrm{sat}}^2/Q^2$ modulo  logarithmic corrections, with energy dependence driven by the
saturation scale. The mild dependence at $\tau_A\simle 1$
corresponds to the fact that the cross section scales as
$\propto \sigma_0\log (Q_{\mathrm{sat}}^2/m_f^2)$ towards  the photoproduction limit. The main features present in the cross sections, which are driven by the scaling function Eq. (\ref{sigtot_param_tau}), can be qualitatively reproduced in the phenomenological saturation models (see discussion in Ref. \cite{magvicprl}). The expressions get simplified to $\sigma_{\mathrm{scal}}^{\mathrm{boson}}\propto \tau_A^{-\gamma}\,[1+\log\,(\tau_A)]$ ($\gamma \simeq 1$). 

\begin{figure}[t]
\includegraphics[scale=0.42]{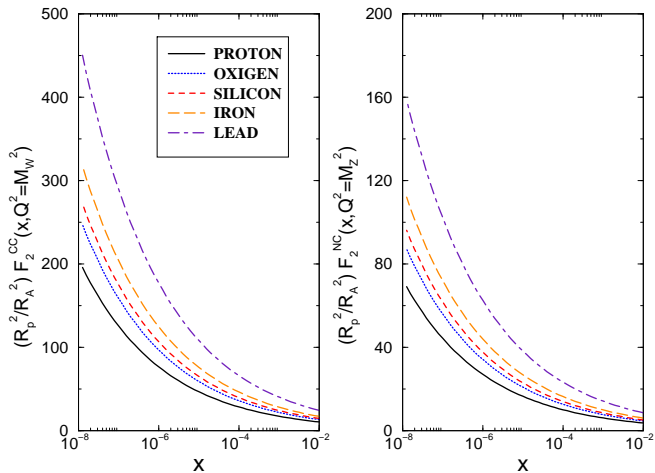}
\caption{The charged (left panel) and neutral (right panel) current structure functions times $R_p^2/R_A^2$  versus Bjorken-$x$ for different atomic nuclei at $Q^2=m_{W}^2 \,(m_Z^2)$.}
\label{fig:2}
\end{figure}

In Fig. \ref{fig:2} one shows the charged and neutral current structure functions, given by Eq. (\ref{weak_sfs}), rescaled by the factor $R_p^2/R_A^2$ as a function of $x$ and distinct nuclei, including the proton case. They are computed at scale $Q^2=m_W^2 \,(m_Z^2)$ and plotted down to extremely small $x\simeq 10^{-8}$. For the proton case, the results are similar to the saturation model, whereas stay a factor about 1.7 smaller than the unified BFKL-DGLAP approach with screening effects and a factor 3 in the case without screening (for more detail on these results, see Refs. \cite{nu_prd,STASTO}). This has consequences in the reduction of the magnitude of the neutrino cross section due to the nonlinear perturbative QCD effects. The saturation effects slow down the rise of the structure functions as $x$ decreases. These effects are sizeable even at larger values of order $Q^2\simeq m_W^2$ because they also contribute to the leading twist part of the structure functions due to the geometric scaling window. Concerning the nuclear structure functions,  the  nuclear effects at the scale $Q^2\simeq m_W^2$ are relatively smaller than the usual expectation $F_2^A \propto A^{1/3}F_2^N$. As in the present calculation $Q_{\mathrm{sat\,,A}}^2 \propto A^{4/9}$, a smaller nuclear shadowing should be expected. Therefore, this implies a smaller reduction in the magnitude of the neutrino-nucleus inelastic scattering due to nuclear shadowing in contrast with recent approaches using the strength of the nonlinear term proportional to $A^{1/3}$.

The total CC (NC) neutrino-nucleon cross sections as a function of the neutrino energy and atomic number are given by the integration over available phase space and read as, 
\begin{eqnarray}
& & \sigma^{\mathrm{CC,NC}}_{(\nu,\,\bar{\nu})}(E_{\nu};\,A)=\int _{Q_{\mathrm{min}}^2}^{s} \! dQ^2\int_{Q^2/s}^1 \! dx \,\frac{1}{xs}
\frac{\partial^2 \,\sigma_{(\nu,\,\bar{\nu})}^{\mathrm{CC,NC}}}{\partial x\,\partial y}, \label{signutotal}\\
& & \frac{\partial^2 \,\sigma_{(\nu,\,\bar{\nu})}^{\mathrm{CC,NC}}}{\partial x\,\partial y} = \frac{G_F^2\,m_N \,E_{\nu}}{\pi}\,\left(\frac{m_{W,Z}^2}{Q^2+m_{W,Z}^2}\right)^2 \times\nonumber\\ 
& &\left[\frac{1+(1-y)^2}{2} F_2^{\mathrm{CC,NC}}(x,Q^2) - \frac{y^2}{2}F_L^{\mathrm{CC,NC}}(x,Q^2)\right],
\label{difxsecnu}
\end{eqnarray}
where a  minimum $Q_{\mathrm{min}}^2 \propto {\cal O}(1)$ GeV$^2$ is introduced in order to stay in the DIS region and $G_F$ is the Fermi constant. Here, one considers UHE neutrinos, where the valence quark contribution stays constant  and physics is driven  by sea quark contributions. Hence, the $xF_3^{\mathrm{CC,NC}}$ contribution should be negligible and it will be disregarded. In Eq. (\ref{difxsecnu}), the nuclear dependence on the structure functions is implicit. 

\begin{figure}[t] 
\begin{tabular}{cc} 
\epsfig{file=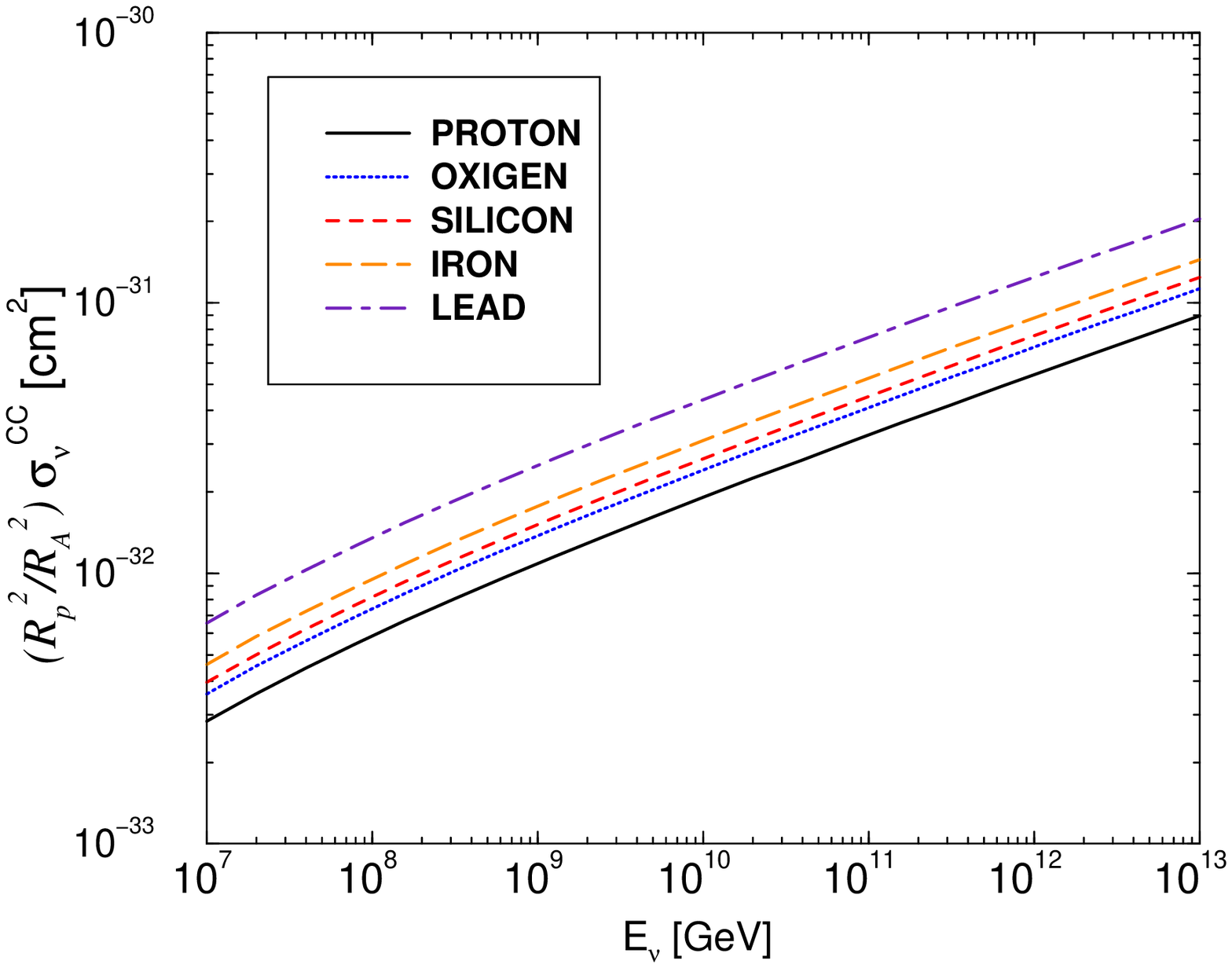,width=43mm,height=55mm} & \epsfig{file=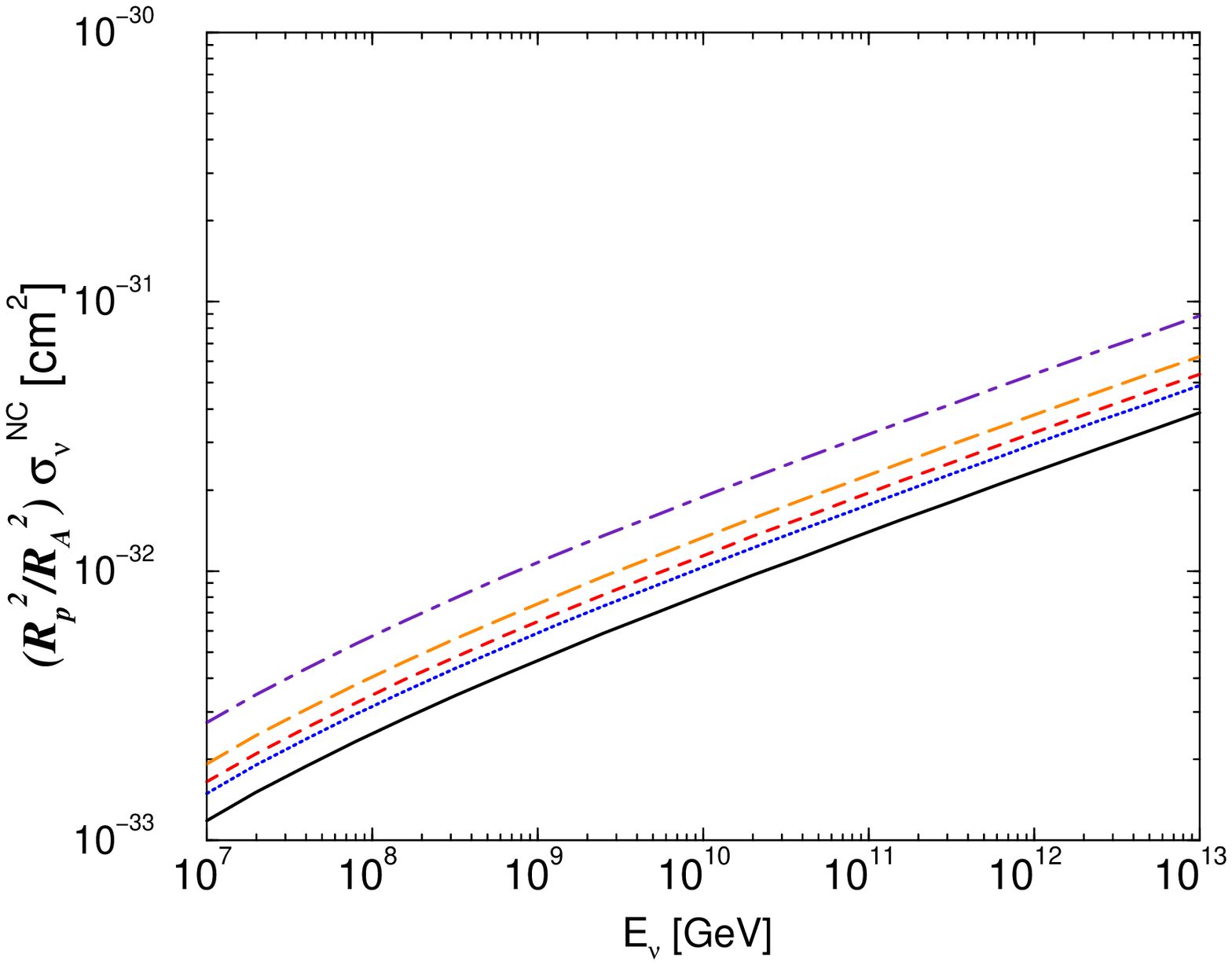,width=43mm,height=55mm}
\end{tabular} 
\caption{The neutrino-nucleus CC (left panel) and NC (right panel) cross sections as a function of neutrino energy $E_{\nu}$ and for different atomic nuclei. The curves are rescaled by the factor $R_p^2/R_A^2$.}
\label{fig:3}
 \end{figure}

\begin{figure}[t]
\includegraphics[scale=0.42]{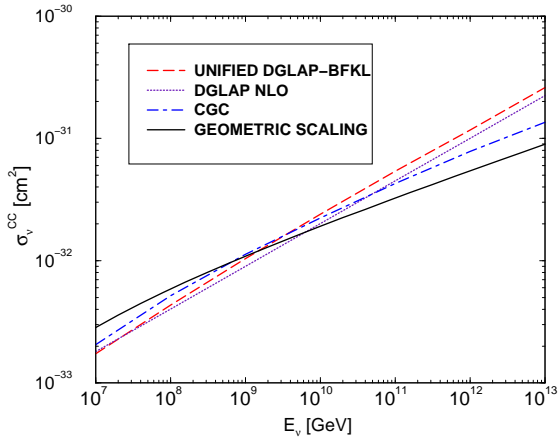}
\caption{Comparison with other approaches: unified DGLAP-BFKL, NLO DGLAP and CGC calculations (see text). Present result (geometric scaling) given by the solid line .}
\label{fig:4}
\end{figure}

In Fig. \ref{fig:3} one shows CC and NC neutrino cross sections, given by Eqs. (\ref{signutotal}), rescaled by the factor $R_p^2/R_A^2$ as a function of neutrino energy $E_{\nu}$ and distinct nuclei, including the nucleon. Based on the color dipole picture, one used the simple relation $F_L \approx (2/11)F_2$ in Eq. (\ref{difxsecnu}), which gives a very small contribution. It is worth mentioning that the calculations are also shown for either low neutrino energies ($E_{\nu}\simle 10^{11}$ GeV), where a part of the contribution to the integrated cross section would be out the scaling window. For the nucleon case, the results are close to the saturation models \cite{nu_prd}. For the nuclear case, the results indicate a weak nuclear shadowing, as discussed above. A comparison with other high energy approaches is presented in Fig. \ref{fig:4} for the nucleon case (see Ref. \cite{nu_prd} for detail on those results). At energies $E_{\nu}\simge 10^{12}$ GeV, the current calculation produces a reduction  by a factor 2 in relation to both NLO DGLAP and unified BFKL-DGLAP approaches and a milder energy dependence. On the other hand, the result has similar behavior as the CGC model. The reason is that the CGC model takes the geometric scaling property and its extrapolation to the saturation region in the dipole cross section.
 
The property of the geometric scaling allows to  analytically calculate the UHE neutrino cross section given by Eq. (\ref{signutotal}). The integration on Bjorken-$x$ can be carried out by making a change of variables. The integration on $x$ is then replaced by an integration on $\tau_A$ of the scaling function, Eq. (\ref{sigtot_param_tau}). This produces the explicit  expression $d\sigma/dQ^2 \propto \beta\,{}_3\!F_3\,[1,1,1;2,2,2;-\beta]_{\tau_{\mathrm{min}}}^{\tau_{\mathrm{max}}}$, where $\tau_{\mathrm{min}}=Q^2/Q_{\mathrm{sat}}^2(x=Q^2/s)$,  $\tau_{\mathrm{max}}=Q^2/Q_{\mathrm{sat}}^2(x=1)$ and ${}_3\!F_3$ is a hypergeometric function. The further integration on $Q^2$ can be performed, using the remaining leading term on virtuality and supposing $y=Q^2/xs \ll 1$. The result using the leading term represents with good accuracy the full numerical calculation. The ratio between the full calculation and the leading term is independent of energy and takes a constant value for $E_{\nu} \simge 10^{8}$ GeV, giving $R_{\mathrm{cor}}=\sigma_{\nu}(\mathrm{exact})/\sigma_{\nu}(\mathrm{approx.})= 0.82$. Therefore, the analytical calculation provides a theoretical parameterization for the UHE neutrino-nucleus cross sections based on the geometric scaling property. They are as follows,
\begin{eqnarray}
\sigma^{\mathrm{CC,\,NC}}_{(\nu,\,\bar{\nu})}  =  {\cal N}_{(i)} \,A^{\alpha}\left(\frac{R_A^2}{R_p^2}\right)^{1-\alpha}\!\left[ C_1^{(i)}\,E_{\nu}^{\,\omega_{\mathrm{scal}}} - C_2^{(i)}\right]\,,
\end{eqnarray}
where ${\cal N}_{(i)}$ are overall normalizations, $C_{1,\,2}^ {(i)}$ are  numerical constants with $i=\mathrm{CC,\,NC}$, $\omega_{\mathrm{scal}}=b\lambda$ and $\alpha = b/\delta$. This implies in a mild power-like rise $\omega_{\mathrm{scal}}\simeq 0.2$ for the neutrino cross section in contrast with other theoretical approaches. The nuclear dependence is approximately linear, $\sigma_{\nu,\bar{\nu}}^{\mathrm{nuclei}}\propto A\,\sigma_{\nu,\bar{\nu}}^{\mathrm{nucleon}}$,  once $b\simeq \delta$ and hence $\alpha \approx 1$.  The remaining constants are given by,
\begin{eqnarray}
& &  {\cal N}_{(i)}   =  R_{\mathrm{cor}}\left(\frac{\bar{\sigma}_0\,G_F^2\,m_{W,Z}^2}{8\pi^3\lambda}\right)\left(\frac{a\,x_0^{\,\omega_{\mathrm{scal}}}}{b}\right)\frac{B_{(i)}}{\alpha_{\mathrm{em}}\,\sum_f e_f^2}\,,\\
& & C_1^{(i)}  = \pi\, (2\,m_N)^{\omega_{\mathrm{scal}}}\left(m_{W,Z}^2\right)^{-\nu_{\mathrm{scal}}}\mathrm{csc}\left(\pi \,\nu_{\mathrm{scal}}\right)\,\left(1-\nu_{\mathrm{scal}} \right)\nonumber ,\\
& & C_2^{(i)} = \pi\, \left(m_{W,Z}^2\right)^{-b}\mathrm{csc}\left(\pi \,b\right)\,\left(1-b \right)\nonumber ,
\end{eqnarray}
where one uses the notation $B_{\mathrm{CC}} = 4$, $B_{\mathrm{NC}} = K_{\mathrm{chiral}}$ and $\nu_{\mathrm{scal}}=b-\omega_{\mathrm{scal}}$. Numerically, this gives a total cross section $\sigma_{\nu,\bar{\nu}}^{tot}= 1.48\times10^{-34}\,A^{\alpha}(E_{\nu}/\mathrm{GeV})^{0.227}$ cm$^2$. The cross section above can have implications for neutrino observatories because experiments are planned to detect UHE by observation of the nearly horizontal air showers in Earth coming from neutrino-air interactions \cite{KUSENKO}. A reduced cross section produces a smaller event rate for such neutrino-induced showers and could compromise the detection signal. However, the rate of up-going air showers initiated by muon and tau leptons produced in neutrino-nucleon reactions just below the surface would increase, being possibly larger than the horizontal air shower rate.
 
In summary, it is demonstrated that the cross sections for $W^{\pm}(Z^0)$-nucleus  processes in UHE inelastic neutrino scattering should exhibit  geometric scaling on the single scaling variable $\tau_A =Q^2/Q_{\mathrm{sat,A}}^2$. This implies that such dimensionless scale  absorbs their energy and atomic number dependences. Based on this property, an analytical calculation at high energies is made possible. This allows to propose a theoretical parameterization for the UHE neutrino cross sections supposed to be valid at energies $E_{\nu} \simge 10^{8}$ GeV. The resulting cross section has a mild energy behavior on energy in comparison with the usual QCD calculations based on linear evolution equations. Moreover, the nuclear dependence is approximately linear on the atomic number.

\vspace{-0.5cm}

\begin{acknowledgments}

\vspace{-0.3cm}

 The author is grateful for the warm hospitality and financial support of Departamento de F\'{\i}sica de Part\'{\i}culas, Universidade de Santiago de Compostela, where this work was accomplished. Special thanks go to Elena Ferreiro and Nestor Armesto for useful remarks and comments on the manuscript. The FORTRAN codes are available at $\tt{http://www.if.ufrgs.br/\,_{\widetilde{}}\,\,magnus/neutrinos.html}$. 

\end{acknowledgments}


\begin{thebibliography}{99}


\bibitem{travwaves} 
S.~Munier and R.~Peschanski,
Phys.\ Rev.\ Lett.\  {\bf 91}, 232001 (2003).


\bibitem{KPP}
R.~A. Fisher,
\newblock Ann. Eugenics {\bf 7}, 355 (1937);
A.~Kolmogorov, I.~Petrovsky, and N.~Piscounov,
\newblock Moscou Univ. Bull. Math. {\bf A1}, 1 (1937).

\bibitem{BK}
I.~Balitski\u{\i},
Nucl.\ Phys.\ {\bf B463}, 99 (1996);
Y.~V. Kovchegov,
\newblock Phys. Rev. D {\bf 60}, 034008 (1999);
\newblock
{\bf 61}, 074018 (2000).

\bibitem{Iancu:2002tr}
E.~Iancu, K.~Itakura and L.~McLerran,
Nucl.\ Phys.\ A {\bf 708}, 327 (2002); A.H. Mueller and D.N. Triantafyllopoulos,  Nucl. Phys. {\bf B640}, 331 (2002).


\bibitem{Stasto:2000er}
A.~M.~Sta\'sto, K.~Golec-Biernat and J.~Kwiecinski,
Phys.\ Rev.\ Lett.\  {\bf 86}, 596 (2001).


\bibitem{Freund:2002ux}
A.~Freund, K.~Rummukainen, H.~Weigert and A.~Schafer,
Phys.\ Rev.\ Lett.\  {\bf 90}, 222002 (2003).


\bibitem{magvicprl}
V.~P.~Goncalves and M.~V.~T.~Machado,
Phys.\ Rev.\ Lett.\  {\bf 91}, 202002 (2003).

\bibitem{gomampla}
V.~P.~Goncalves and M.~V.~T.~Machado,
Mod. Phys. Lett. A {\bf 19}, 2525  (2004).

\bibitem{Armesto_scal}
N.~Armesto, C.~A.~Salgado and U.~A.~Wiedemann, Phys.\ Rev.\ Lett.\  {\bf 94}, 022002 (2005).
\bibitem{GBW} K. Golec-Biernat and  M. W\"usthoff,  Phys. Rev. D {\bf 59}, 014017 (1999),  {\it ibid.} {\bf 60} 114023 (1999).


\bibitem{nu_prd}
M.~V.~T.~Machado, Phys.\ Rev.\ D {\bf 70}, 053008 (2004).

\bibitem{KUTAK} K. Kutak and J. Kwieci\'nski,  Eur. Phys. J. C {\bf 29}, 521 (2003). 

\bibitem{DIPOLEPIC}
A. H. Mueller,  Nucl. Phys. {\bf B335}, 115 (1990);
N.N. Nikolaev and B.G. Zakharov,  Z. Phys. {\bf C49}, 607 (1991). 

\bibitem{JAMAL} J. Jalilian-Marian,  Phys. Rev. D {\bf 68}, 054005 (2003).

\bibitem{STASTO} A.M. Sta\'sto, Int.\ J.\ Mod.\ Phys.\ A {\bf 19}, 317 (2004).

\bibitem{KUSENKO} A. Kusenko and T.J. Weiler,  Phys. Rev. Lett. {\bf 88}, 161101(2002).



\end{thebibliography}
\end{document}